\documentclass[conference]{IEEEtran}
\IEEEoverridecommandlockouts
\usepackage{cite}
\usepackage{amsmath,amssymb,amsfonts}
\usepackage{bm}
\usepackage{algorithmic}
\usepackage{graphicx}
\usepackage{textcomp}
\usepackage{xcolor}

\def\BibTeX{{\rm B\kern-.05em{\sc i\kern-.025em b}\kern-.08em
    T\kern-.1667em\lower.7ex\hbox{E}\kern-.125emX}}
    
\begin{document}

\title{EffiSegNet: Gastrointestinal Polyp Segmentation through a Pre-Trained EfficientNet-based Network with a Simplified Decoder  \\

\thanks{Funded by the European Union (DIOPTRA, 101096649). Views and
opinions expressed are, however, those of the author(s) only and do not
necessarily reflect those of the European Union or the Health and Digital
Executive Agency. Neither the European Union nor the granting authority can
be held responsible for them. This work has received funding from the Swiss
State Secretariat for Education, Research and Innovation (SERI). Funded by
UK Research and Innovation (UKRI) under the UK government’s Horizon
Europe funding guarantee [grant number 10056682].}
}

\author{\IEEEauthorblockN{Ioannis A. Vezakis}
\IEEEauthorblockA{\textit{TECREANDO B.V.} \\
Amsterdam, The Netherlands \\
0000-0003-4976-4901}
\and
\IEEEauthorblockN{Konstantinos Georgas}
\IEEEauthorblockA{\textit{Biomedical Engineering Laboratory} \\
\textit{School of Electrical and Computer Engineering}\\
\textit{National Technical University of Athens}\\
Athens, Greece \\
0000-0002-2832-3747}
\and
\IEEEauthorblockN{Dimitrios Fotiadis}
\IEEEauthorblockA{\textit{Dept. of Materials Science and Engineering} \\
\textit{University of Ioannina}\\
Ioannina, Greece \\
0000-0002-7362-5082}
\and
\IEEEauthorblockN{George K. Matsopoulos}
\IEEEauthorblockA{\textit{Biomedical Engineering Laboratory} \\
\textit{School of Electrical and Computer Engineering}\\
\textit{National Technical University of Athens}\\
Athens, Greece \\
0000-0002-2600-9914}
}

\maketitle

\begin{abstract}
This work introduces EffiSegNet, a novel segmentation framework leveraging transfer learning with a pre-trained 
Convolutional Neural Network (CNN) classifier as its backbone. Deviating from traditional architectures with a symmetric 
U-shape, EffiSegNet simplifies the decoder and utilizes full-scale feature fusion to minimize computational cost and the 
number of parameters. We evaluated our model on the gastrointestinal polyp segmentation task using the publicly 
available Kvasir-SEG dataset, achieving state-of-the-art results. Specifically, the EffiSegNet-B4 network variant achieved 
an $F$1 score of 0.9552, mean Dice (mDice) 0.9483, mean Intersection over Union (mIoU) 0.9056, Precision 0.9679, 
and Recall 0.9429 with a pre-trained backbone -- to the best of our knowledge, the highest reported scores in the literature 
for this dataset. Additional training from scratch also demonstrated exceptional performance compared to previous work, 
achieving an $F$1 score of 0.9286, mDice 0.9207, mIoU 0.8668, Precision 0.9311 and Recall 0.9262. These results 
underscore the importance of a well-designed encoder in image segmentation networks and the effectiveness of 
transfer learning approaches.
\end{abstract}

\begin{IEEEkeywords}
medical images, colonoscopy, endoscopy, polyp segmentation, semantic segmentation, convolutional neural networks, transfer learning, efficientnet
\end{IEEEkeywords}

\section{Introduction}
Colorectal Cancer (CRC) is one of the most prevalent cancers in Europe, accounting for 12.9\% of all new 
cancer diagnoses and 12.4\% of deaths in 2022 \cite{ECISEuropeanCancer}.
Colonoscopy is the current gold standard in the early detection and diagnosis of colorectal abnormalities,
particularly in the identification of colon polyps, a potential precursor to CRC \cite{jhaResUNetAdvancedArchitecture2019}.
As medical imaging technologies advance, there is a growing demand for accurate and efficient tools to 
assist clinicians in polyp detection, as miss rates with the current manual approach are estimated
between 14-30\% depending on the polyp type and size \cite{jhaResUNetAdvancedArchitecture2019}.

In recent years, deep learning approaches have demonstrated remarkable success in various medical 
image analysis tasks, leveraging large datasets and pre-trained models to achieve state-of-the-art results. 
Transfer learning, in particular, has emerged as a promising technique to address the data scarcity issue, 
allowing models trained on external datasets to adapt and excel in specific medical imaging domains \cite{kimTransferLearningMedical2022}.

Despite the effectiveness of transfer learning, the predominant methodologies employed for colon polyp 
segmentation, as exemplified by widely-used networks like U-Net \cite{ronnebergerUNetConvolutionalNetworks2015}, ResUNet 
\cite{jhaKvasirSEGSegmentedPolyp2020}, and ResUNet++ \cite{jhaResUNetAdvancedArchitecture2019}, 
often opt for training from scratch on 
the task-specific dataset. Notable exceptions, mainly involving transformer networks, still fall short when 
compared to the current best performing network, DUCK-Net, a Convolutional Neural Network (CNN)
trained from scratch \cite{dumitruUsingDUCKNetPolyp2023}.
This paradox is the main motivation behind revisiting transfer learning techniques for segmentation networks.
Current approaches usually employ symmetric U-shaped networks, with the encoder consisting of a pre-trained CNN
classifier, and the decoder a symmetric stack of convolutional layers with randomly initialized weights, that refine
concatenated feature maps from previous layers \cite{iglovikovTernausNetUNetVGG112018, kalininMedicalImageSegmentation2020, sharmaLiSegPNetEncoderDecoderMode2023, bal-ghaouiUNetTransferLearning2023}.

Recently, Lu \textit{et al.} \cite{luHalfUNetSimplifiedUNet2022} demonstrated that the divide-and-conquer strategy in the encoder
of the U-Net is the main contributor to its effectiveness. In their work, they designed Half-UNet, a segmentation
network that does not have the typical symmetric U-shape. Instead, their network is simplified by utilizing full-scale
fusion of the encoder's outputs, and refinement using two Ghost modules. This design achieved superior
segmentation efficiency in terms of computational cost while maintaining comparable accuracy.

Driven by these considerations, we propose a novel network architecture named EffiSegNet, which 
deviates from previous transfer learning approaches by utilizing the EfficientNet family of CNNs \cite{tanEfficientNetRethinkingModel2019} 
as the encoder, and discarding the symmetric U-shape for a simplified decoder that keeps the number of added 
parameters and 
complexity to a minimum. We demonstrate EffiSegNet's effectiveness on Kvasir-SEG, a gastrointestinal polyp 
segmentation dataset, where its performance surpasses current state-of-the-art models. 
To ensure the reproducibility of our research, our code and dataset splits are publicly available on Zenodo 
(https://doi.org/10.5281/zenodo.10601024).

\section{Network Architecture}

Inspired by Half-UNet's simplified U-Net architecture, we utilized the EfficientNet family 
of CNN classifiers \cite{tanEfficientNetRethinkingModel2019} as the backbone to create several variants of a 
new network architecture which we named EffiSegNet.
The core of the network is comprised of an EfficientNet CNN pre-trained on the ImageNet
classification dataset. The overall network's architecture has been intentionally designed to minimize the
amount of non pre-trained parameters, thereby reducing the corresponding computational overhead
and the number of randomly initialized weights.
Using EfficientNet as the encoder of the network, the final feature maps produced before each
downsampling step are extracted. In a typical U-Net architecture, these feature maps are upsampled,
concatenated with the features of the previous stage along the channel dimension, and then refined 
using consecutive convolutional layers:
\begin{equation}
	\label{u-net_concat}
	\tilde {\bm{x}}_s = F_s(concat(\bm{x}_s, up(\bm{x}_{s+1}))).
\end{equation}
In this context, the stage $s$ refers to a distinct level in the network where the spatial dimensions of 
the feature maps (i.e. their height $H$ and width $W$) are reduced by a factor of 2. 
Therefore, $\bm{x}_s$ denotes the output feature maps of the stage $s$, and $\bm{x}_{s+1}$ the output of
the subsequent stage with the spatial dimensions halved. $F_s(\cdot)$ is the stack of convolutional 
layers that refine the fused features, $up(\cdot)$ is the upsampling operation that doubles the 
spatial size, and $concat(\cdot, \cdot)$ is the concatenation operation between two stacks of feature maps.

The feature fusion method described in Eq. \ref{u-net_concat}, although effective, results in
memory and computational overhead, which previous work has avoided by performing element-wise 
addition instead \cite{He_2016_CVPR, luHalfUNetSimplifiedUNet2022}. However, for the addition operation to be
peformed, the feature maps need to match across all dimensions (height, width, number of channels). 
To this end, we employ first a simple convolutional layer, followed
by batch normalization, and upsampling using nearest-neighbor interpolation, in order to equalize 
the dimensions across all stages. The optimal number of channels was heuristically found to be 32.
We opted to perform the convolution operation before the upsampling due to the reduced computational
complexity. This operation is defined as:
\begin{equation}
	\tilde {\bm{x}}_s = \sum _{s=1} ^ n up(F_s(\bm{x}_s))+ F_0(\bm{x}_0),
\end{equation}
where $n$ is the network depth (equal to 5 for EfficientNets), $F_s(\cdot)$ is the simple convolutional layer
that outputs 32 feature maps, followed by batch normalization, and $up(\cdot)$ is the upsampling operation 
that increases the spatial dimensions to those of the original input's size, instead of doubling them.

\begin{figure}[htbp]
\centerline{\includegraphics[width=0.5\textwidth]{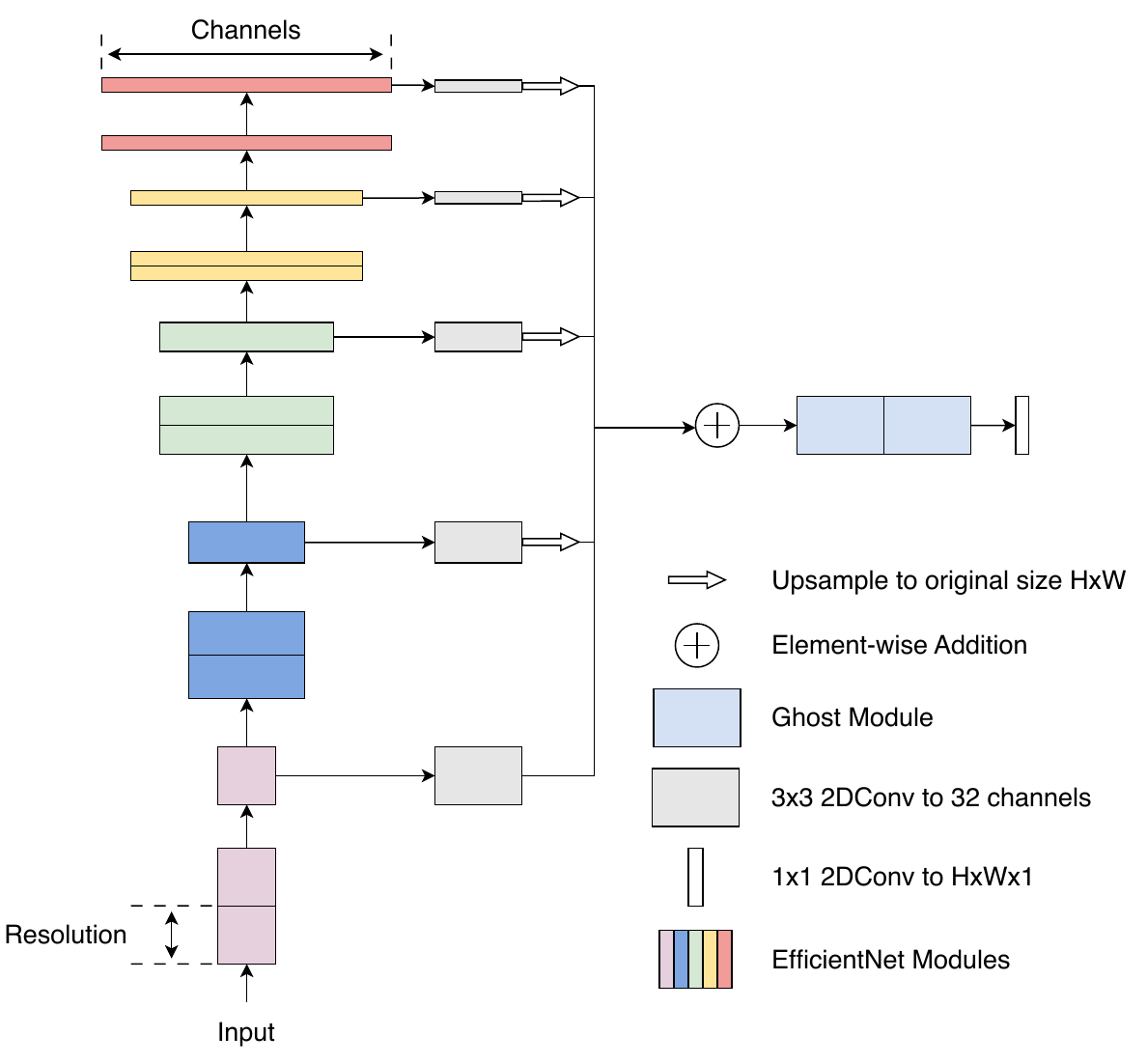}}
\caption{The EffiSegNet architecture. A pre-trained EfficientNet model serves as the backbone of the network, scaling it up and down using compound scaling.}
\label{effisegnet_architecture}
\end{figure}

Following feature fusion across all stages, two Ghost modules \cite{hanGhostNetMoreFeatures2020} are utilized, 
a strategy similarly employed in \cite{luHalfUNetSimplifiedUNet2022}. These modules effectively generate more feature maps 
using a limited number of parameters and operations, thereby contributing to the reduction of non pre-trained parameters.
The final layer involves a simple 1$\times$1 convolution, followed by a sigmoid activation function which produces 
the final output.
The network's architecture, illustrated in Fig. \ref{effisegnet_architecture}, can be easily and efficiently scaled up or 
down in terms of its depth, width, and resolution, by utilizing EfficientNet's compound scaling technique 
\cite{tanEfficientNetRethinkingModel2019}.
Following a similar naming scheme to the original EfficientNet implementation, we named each scaled version of our 
network as ``EffiSegNet-BN'', where \textit{N} corresponds to the EfficientNet variant used as the backbone.
Table \ref{number_of_params} depicts the number of pre-trained and randomly initialized parameters for each 
network variant.

\begin{table}[htbp]
\caption{Number of Pre-trained and Randomly Initialized Parameters}
\begin{center}
\begin{tabular}{|c|c|c|c|}
\hline
\textbf{Network} & \textbf{Pre-trained}& \textbf{Randomly}& \textbf{Random to} \\
\textbf{Variant} & \textbf{Params}& \textbf{Init. Params}& \textbf{Pre-trained Ratio} \\
\hline
EffiSegNet-B0 & 4.0M & 0.15M & 3.8\% \\
\hline
EffiSegNet-B1 & 6.5M & 0.15M & 2.3\% \\
\hline
EffiSegNet-B2 & 7.7M & 0.16M & 2.1\% \\
\hline
EffiSegNet-B3 & 10.7M & 0.18M & 1.7\% \\
\hline
EffiSegNet-B4 & 17.5M & 0.21M & 1.2\% \\
\hline
EffiSegNet-B5 & 28.3M & 0.24M & 0.8\% \\
\hline
EffiSegNet-B6 & 40.7M & 0.27M & 0.7\% \\
\hline
EffiSegNet-B7 & 63.8M & 0.3M & 0.5\% \\
\hline
\end{tabular}
\label{number_of_params}
\end{center}
\end{table}

\section{Experimental Setup}

We tested the EffiSegNet variants on Kvasir-SEG \cite{jhaKvasirSEGSegmentedPolyp2020}, an open-access 
segmentation dataset containing 1000 endoscopic images of gastrointestinal polyps and their corresponding ground truth
delineations.
To ensure an equal comparison with current state-of-the-art, we used the 80:10:10 split into
training, validation, and testing subsets provided by Dumitru \textit{et al.} (2023) \cite{dumitruUsingDUCKNetPolyp2023}.
To the best of our knowledge, their approach using the DUCK-Net architecture is, until
now, the best performing approach on this particular dataset.

We trained all EffiSegNet variants using a batch size of 8 for 300 epochs. In cases where the
available memory was insufficient, the maximum possible batch size was determined by performing
a binary search. The Adam optimizer with decoupled weight decay regularization was used 
\cite{loshchilovDecoupledWeightDecay2019}, with an initial learning rate of $10^{-4}$. This was gradually reduced
to $10^{-5}$ over the course of training using cosine annealing of the learning rate.
The loss function used was the average of the Dice and Cross Entropy loss.

The spatial resolution of the original input images varied between 332$\times$487 to 1920$\times$1072
pixels. These images were resized using Lanczos interpolation to the spatial dimensions on which each 
particular EfficientNet variant
was pre-trained on. This is 224$\times$224 for EfficentNetB0, 240$\times$240 for EfficientNetB1, 260$\times$260
for EfficientNetB2, 300$\times$300 for EfficientNetB3, 380$\times$380 for EfficientNetB4, 456$\times$456
for EfficientNetB5, 528$\times$528 for EfficientNetB6, and 600$\times$600 for EfficientNetB7.
Previous work has suggested that pre-trained EfficientNets work best on images with similar dimensions
to those they were pre-trained on \cite{vezakisDeepLearningApproaches2023}, therefore, we did not opt to resize to alternative
dimensions.

We followed the augmentation techniques used in \cite{dumitruUsingDUCKNetPolyp2023}, with the addition of elastic
deformation. More specifically, during training we applied: 
\begin{itemize}
	\item Random horizontal and vertical flip.
	\item Color jitter with the brightness chosen randomly between 0.6 and 1.6, a contrast factor of 0.2, saturation factor 0.1 and hue factor 0.01.
	\item Affine transformation with scale value uniformly sampled between 0.5 and 1.5, translation up to 12.5\% of the image height and width, and rotation between -90 and 90 degrees.
	\item Elastic deformation with the Gaussian filter sigma set to 50, alpha value of 1, and Lanczos interpolation.
\end{itemize}
Finally, all images were normalized using the channel mean and standard deviation of ImageNet: [0.485, 0.456, 0.406] and [0.229, 0.224, 0.225] for each of RGB channels, respectively.

\section{Results}

Table \ref{results} depicts the results as measured for the test subset.
We have computed the $F$1 Score, mean Dice, mean Intersection over Union (IoU), Precision and Recall for all our network variants.
However, not all of the metrics were reported in all works (n/a -- not available flag).

\begin{table}[htbp]
\caption{Segmentation Results on the Kvasir-SEG dataset.}
\begin{center}
\begin{tabular}{|c|c|c|c|c|c|}
\hline
\textbf{Model} & \textbf{$F$1 Sc.}& \textbf{mDice}& \textbf{mIoU}& \textbf{Precision}& \textbf{Recall} \\
\hline
U-Net$^ \dag$ \cite{ronnebergerUNetConvolutionalNetworks2015} & 0.8655 & n/a & 0.7629 & 0.8593 & 0.8718 \\
\hline
ResUNet \cite{jhaKvasirSEGSegmentedPolyp2020} & 0.7878 & n/a & 0.7778 & n/a & n/a \\
\hline
ResUNet++ \cite{jhaResUNetAdvancedArchitecture2019} & 0.8133 & n/a & 0.7927 & 0.7064 & 0.8774 \\
\hline
Li-SegPNet* \cite{sharmaLiSegPNetEncoderDecoderMode2023} & 0.9058 & n/a & 0.8800 & 0.9424 & 0.9254 \\
\hline
PraNet*$^ \dag$ \cite{fanPraNetParallelReverse2020} & 0.9094 & n/a & 0.8339 & 0.9599 & 0.8640 \\
\hline
ColonFormer* \cite{ducColonFormerEfficientTransformer2022} & n/a & 0.927 & 0.877 & n/a & n/a \\
\hline
DUCK-Net \cite{dumitruUsingDUCKNetPolyp2023} & 0.9502 & n/a & 0.9051 & 0.9628 & 0.9379 \\
\hline
EffiSegNet-B0* & 0.9421 & 0.9304 & 0.8794 & 0.9475 & 0.9368 \\
\hline
EffiSegNet-B1* & 0.9448 & 0.9288 & 0.8784 & 0.9437 & 0.9461 \\
\hline
EffiSegNet-B2* & 0.9464 & 0.9329 & 0.8836 & 0.9550 & 0.9380 \\
\hline
EffiSegNet-B3* & 0.9465 & 0.9358 & 0.8876 & 0.9613 & 0.9321 \\
\hline
EffiSegNet-B4* & \textbf{0.9552} & 0.9483 & 0.9056 & 0.9679 & \textbf{0.9429} \\
\hline
EffiSegNet-B5* & 0.9513 & \textbf{0.9488} & \textbf{0.9065} & 0.9713 & 0.9321 \\
\hline
EffiSegNet-B6* & 0.9531 & 0.9477 & 0.9060 & \textbf{0.9724} & 0.9334 \\
\hline
EffiSegNet-B7* & 0.8289 & 0.7629 & 0.7073 & 0.8957 & 0.7713 \\
\hline
\multicolumn{6}{l}{$^*$ Model was pre-trained on an external dataset.}\\
\multicolumn{6}{l}{$^ \dag$ Evaluation scores from \cite{dumitruUsingDUCKNetPolyp2023}.}
\end{tabular}
\label{results}
\end{center}
\end{table}

\section{Training from Scratch}

We conducted a separate experiment with EffiSegNet-B4, the best performing network variant in terms of the
$F$1 score, and re-trained it with randomly initialized weights to determine the pre-training's effect on the network's
performance. The results of this experiment are reported in Table \ref{ablation}.

\begin{table}[htbp]
\caption{Comparison of Pre-Trained and Randomly Initialized Network Performance}
\begin{center}
\begin{tabular}{|c|c|c|c|c|c|}
\hline
\textbf{Model} & \textbf{$F$1 Score}& \textbf{mDice}& \textbf{mIoU}& \textbf{Precision}& \textbf{Recall} \\
\hline
EffiSegNet-B4* & 0.9552 & 0.9483 & 0.9056 & 0.9679 & 0.9429 \\
\hline
EffiSegNet-B4 & 0.9286 & 0.9207 & 0.8668 & 0.9311 & 0.9262 \\
\hline
\multicolumn{6}{l}{$^*$ Model was pre-trained on an external dataset.}\\
\end{tabular}
\label{ablation}
\end{center}
\end{table}

\section{Discussion}

The effectiveness of transfer learning in improving medical image analysis on limited data has been 
consistently demonstrated in previous studies \cite{kimTransferLearningMedical2022, koraTransferLearningTechniques2022}.
Yet, the predominant and baseline networks trained on the Kvasir-SEG dataset did not utilize pre-training on any external
datasets \cite{ronnebergerUNetConvolutionalNetworks2015, jhaKvasirSEGSegmentedPolyp2020, jhaResUNetAdvancedArchitecture2019, dumitruUsingDUCKNetPolyp2023}.
Even exceptions to this practice, mainly involving transformer networks, still fall short in performance compared 
to DUCK-Net, a CNN trained from scratch \cite{dumitruUsingDUCKNetPolyp2023}.

In this work, we proposed a novel segmentation network, EffiSegNet, incorporating a pre-trained classifier 
backbone and a minimal number of parameters added on top to transition into pixel-level classification. 
This approach stems from the observation that the encoder's divide-and-conquer strategy outweighs the 
decoder's feature fusion significance, thereby meaning that a symmetric U-shaped network is not 
necessarily optimal \cite{luHalfUNetSimplifiedUNet2022}.

Our results demonstrate that a pre-trained CNN is hard to beat. Specifically, our ``EffiSegNet'' architecture 
achieved state-of-the-art results on the Kvasir-SEG dataset, with larger network variants, namely EffiSegNet-B4,
EffiSegNet-B5, and EffiSegNet-B6, outperforming the current state-of-the-art DUCK-Net in terms of the $F$1 score,
mean IoU, and Precision. 
However, the largest variant, EffiSegNet-B7, was found to be an exception as the overly large amount of 
parameters, detailed in Table \ref{number_of_params}, led to overfitting on the training data. This highlights
the need for careful consideration of model complexity when training on limited datasets.

Training EffiSegNet-B4 without any pre-training resulted in inferior results when compared to its pre-trained
counterpart ($F$1 Score of 0.9286 vs 0.9552), but still among the highest performing networks in the literature.
This further supports that a well designed encoder is much more important than the decoder,
and features from the various stages can be effectively used for pixel-level classification with cheap operations
and few trainable parameters.

Future research could investigate the impact of each stage's features on the final segmentation
accuracy, and explore specialized blocks that better capture features at each scale. Moreover,
given that the EffiSegNet architecture can incorporate any CNN classifier as its backbone, experimentation with different
backbone models could offer new insights.

\section{Conclusion}

This study introduces EffiSegNet, a novel approach to gastrointestinal polyp segmentation on endoscopy images
leveraging transfer learning
and the EfficientNet family of CNNs as the model's backbone. Our findings on the Kvasir-SEG dataset demonstrate
superior performance compared to existing methods, highlighting the effectiveness of incorporating pre-trained networks 
in the model's architecture. The high performance achieved with and without pre-training further 
underscores the significance of prioritizing encoder design over decoder complexity. 
EffiSegNet also provides a versatile framework for integrating any CNN classifier,
opening avenues for future investigation into the impact of different backbone designs. 
As demonstrated by our results, in the evolving field of medical image analysis, EffiSegNet can prove a useful tool 
for enhancing colorectal cancer screening and advancing the application of machine learning in healthcare.

\bibliographystyle{IEEEtran}
\bibliography{IEEEabrv,bibliography.bib}

\end{document}